\documentclass[12pt]{iopart}
\usepackage{graphicx}
\usepackage{color}
\usepackage[colorlinks]{hyperref}

\begin{document}

\title[Simulating quantum transport for a quasi-one-dimensional Bose gas]{Simulating quantum transport for a quasi-one-dimensional Bose gas in an optical lattice: the choice of fluctuation modes in the truncated Wigner approximation}

\author{Bo Xiong$^{1,\dagger}$, Tao Yang$^{1,2,*}$ and Keith A Benedict$^{1}$}
\address{1. Condensed Matter Theory, School of Physics and Astronomy, University of Nottingham, Nottingham NG7 2RD, United Kingdom}
\address{2. Institute of Modern Physics, Northwest University, Xi'an 710069, P. R. China\\
$^\dagger$ Present Address: Seoul National University, Department of Physics and Astronomy, centre for Theoretical Physics, 151-747 Seoul, Korea\\
$^*$ E-mail: yangt@nwu.edu.cn}

\begin{abstract}
We study the effect of quantum fluctuations on the dynamics of a quasi-one-dimensional Bose gas in an optical lattice at zero-temperature using the truncated Wigner approximation with a variety of basis sets for the initial fluctuation modes. The initial spatial distributions of the quantum fluctuations are very different when using a limited number of plane-wave (PW), simple-harmonic-oscillator (SHO) and self-consistently determined Bogoliubov (SCB) modes. The short-time transport properties of the Bose gas, characterized by the phase coherence in the PW basis are distinct from those gained using the SHO and SCB basis. The calculations using the SCB modes predict greater phase decoherence and stronger number fluctuations than the other choices. Furthermore, we observe that the use of PW modes overestimates the extent to which atoms are expelled from the core of the cloud, while the use of the other modes only breaks the cloud structure slightly which is in agreement with the experimental observations \cite{PRL.94.120403}.
\end{abstract}

\maketitle

\section{Introduction}
While the majority of early experiments on Bose-condensed systems of ultra-cold atoms could be explained adequately in terms of the mean-field Gross-Pitaevskii (GP) equation \cite{gp1, gp2}, more recent experiments have probed regimes in which fluctuation effects are either significant or completely invalidate the mean-field treatment. Examples of the former include the damping of the dipole oscillations of a cold-atom cloud in a shallow optical lattice \cite{PRL.94.120403} while the latter include the observation of the superfluid-Mott insulator transition \cite{PRL.81.3108, Nature.415.39}. Quite generally, one expects that in out-of-equilibrium systems, quantum fluctuations can lead to strong dissipation and dynamical instabilities\cite{PRL.93.070401, PRL.95.110403,PRL.95.020402,PRA.72.011601,PRA.73.013605,PRA.74.063625}.

The truncated Wigner approximation (TWA), originally developed in the field of quantum optics, provides an attractive framework to explore the effect of the quantum fluctuations on the dynamics of Bose-Einstein condensates (BECs) \cite{PRL.87.160402,twa2,PRL.94.040401,PRL.95.110403,PRL.100.100402,PRA.72.011601,PRA.74.063625,PRA.58.4824,JLTP.124.431,PRA.66.051602,PRA.73.043617, PRA.74.053605,PRA.79.063624}. The method works by formulating the bosonic many-body problem in the Wigner representation and to note that, when some plausibly small higher order terms are dropped, the Wigner quasi-distribution function satisfies a Fokker-Planck equation. Within this approximation, one can then obtain all correlation functions by simulating the corresponding classical stochastic process. The beauty of the truncated Wigner scheme is that the relevant classical process is governed by the deterministic Gross-Pitaevskii equation with stochastic initial conditions \cite{PRL.94.040401,AP.57.363,PRA.73.043617}
\begin{equation}\label{5.26}
    i\hbar\partial_t\Psi (\mathbf{x},t) =H_{0}\Psi(\mathbf{x},t)+g|\Psi(\mathbf{x},t)|^2\Psi(\mathbf{x},t),
\end{equation}
where $H_{0}= -\hbar^{2}\partial_{\mathbf{x}}^{2}/2m+ V(\mathbf{x},t)$.
The stochastic initial condition is
\begin{equation}
\Psi\left(\mathbf{x},t=0\right) = \psi_0\left(\mathbf{x}\right)  + \xi\left(\mathbf{x}\right)
\end{equation}
where $\psi_0(\mathbf{x})$ is the desired initial form for the mean-field order parameter (such as the form which minimizes the mean field energy functional, found by imaginary time evolution from a simple initial state) and $\xi(\mathbf{x})$ is  a complex gaussian random field:
\begin{eqnarray}
\left\langle \xi( \mathbf{x}) \right \rangle & = & \left\langle \xi^*( \mathbf{x} \right\rangle = 0\\
\left\langle \xi( \mathbf{x}) \xi(\mathbf{x'} \right\rangle & = & \left\langle \xi^*( \mathbf{x}) \xi^*( \mathbf{x'}) \right\rangle = 0\\
\left\langle \xi^*( \mathbf{x}) \xi(\mathbf{x'}) \right\rangle & = & \frac{1}{2} \delta \left( \mathbf{x} - \mathbf{x'} \right) \qquad.
\end{eqnarray}
It is this random field that introduces quantum (and if desired thermal) fluctuations around the mean field behaviour.

In practice one implements this initial condition by selecting an appropriate complete set of basis functions $\left\{\phi_j(\mathbf{x})\right\}$ such that the random field can be written as
\begin{equation}
 \xi\left(\mathbf{x}\right) = \sum_j \xi_j \phi_j(\mathbf{x})
\end{equation}
where the $\xi_j$'s are complex gaussian random variables with
\begin{eqnarray}
\left\langle \xi_j\right\rangle = \left\langle \xi^*_j\right\rangle & = & 0 \\
\left\langle \xi_j\xi_{j'}\right\rangle = \left\langle \xi^*_j\xi^*_{j'}\right\rangle & = & 0\\
\left\langle \xi^*_j\xi_{j'}\right\rangle & = &\frac{1}{2}\delta_{jj'}\qquad.
\end{eqnarray}

Provided a complete set of modes is used for the expansion of quantum fluctuations, the choice of basis would be
immaterial from the point of view of the validity of the method, and the computationally simplest basis preferred.
However, this expansion requires a limited modes to be used, which distinguishes ``low energy'' modes, that properly
contribute to fluctuations in the condensate, from ``high energy'' modes that do not. The use of a truncated basis for the expansion of fluctuations means that it is no longer clear that all basis sets are equivalent. Therefore the dynamics associated with these different initial states is expected to be different. A number of works have used
the TWA with a set of long-wavelength plane-wave (PW) modes to simulate the collision of cold-atom clouds with one
another\cite{PRL.94.040401,PRA.73.043617} and with a surface \cite{PRA.74.053605}. %
One can also apply the Bogoliubov approximation within the framework of the TWA, which is widely used to include fluctuation effects in the study of dynamical properties of BEC systems and will be considered in this paper, as described in Refs.\cite{twa2,PRL.93.070401,PRL.95.110403,PRA.72.011601,PRA.74.063625}. Naturally one supposes that Bogoliubov modes, being well adapted to the mean-field initial state, should give the best account of the physics of any given system but the computational cost of finding these is high and has to be done for each choice of initial state. PW modes, while being straightforward to generate (especially if fast Fourier transform (FFT) based methods are used to solve the GP equation), are poorly adapted to the spatially non-uniform structure of harmonically confined atom clouds. One of our aims is to consider the use of a basis set that is computationally cheaper than the self-consistent determined Bogoliubov (SCB) basis but better adapted to initial states that are harmonically confined condensates.

In this paper we will consider a condensate cloud initially prepared such that $\psi_0(\mathbf{x})$ is chosen to minimize the mean field energy in a harmonic trap. We will then add to this initial fluctuations using three distinct sets of fluctuation modes: PW modes, simple-harmonic-oscillator (SHO) eigenstates and SCB modes. We then follow the evolution of each ensemble in the whole coordinate space when the centre of the harmonic trap is suddenly shifted to one side and a weak optical lattice potential is turned on. While we find, as shown in \cite{PRL.94.120403,PRL.95.110403,PRA.72.011601,PRA.74.063625}, that the inclusion of quantum fluctuations does indeed lead to damping of the dipole oscillations of the centre-of-mass (c.m.) of the condensate, we note that the choice of restricted basis set for the added quantum fluctuations in the initial time has a qualitative effect on the evolution of the spatial density distribution of the cloud, the loss of phase coherence, and the number fluctuations. These qualitative differences originate from the initial distribution of the quantum fluctuations. In the PW basis, some atoms are kicked out from the core part of atomic cloud, resulting in a density distribution with a long tail. This does not agree with the experimental observations \cite{PRL.94.120403}, where the cloud in damped transport has a similar width to the undamped case. However, there is only slight disruption of the cloud structure in the other two cases. Moreover, by using the same number of modes, Bogoliubov modes offer stronger damping of the c.m. trajectory than the others.

The remainder of this paper is organized as follows. In section \ref{model} we summarize the truncated Wigner scheme and our specific implementation of it for the choices of quantum fluctuations in three basis states. In section \ref{disc} we describe the results of the simulations and highlight the differences in the dynamical evolution due to the initial choices of quantum fluctuations. In section \ref{conc} we summarize these and draw more general conclusions about such simulations.

\section{Simulation schemes and numerical methods}\label{model}

We aim to carry out simulations of the situation in the experiment \cite{PRL.94.120403}, in which damped dipole oscillations of a one-dimensional (1D) Bose gas in a shallow optical lattice (OL) were observed. In this experiment, the 1D Bose gas is formed in an array of independent ``tubes'', produced by applying a strong transverse two-dimensional (2D) OL potential to confine a trapped three-dimensional atomic condensate. Then the tubes are corrugated adiabatically by using a very shallow 1D lattice along the axial direction. The dipole oscillations of atoms along the weak axial lattice were excited by suddenly displacing the harmonic trap and the c.m. velocity was imaged.

The transport dynamics of the system based on the above model is studied by using the TWA method. In a very shallow OL, the dynamics of the system can be described by the 1D GP equation (equation \eref{5.26}) with the coupling constant $g=g_{1D}=2\hbar\omega_{\bot}a$. We use the s-wave scattering length $a=5.4$nm for $^{87}Rb$ and $\omega_{\bot}=2\pi\times38$kHz, obtained from the quadratic expansion of the strong transverse 2D OL around the local minima \cite{PRL.94.120403,PRL.95.110403,PRA.72.011601,PRA.74.063625}. The potential energy profile of the 1D optical lattice in the axial direction is characterized by $V_{OL}(x,t)= A(t)\textrm{sin}^2(\pi x/d)$, where $d=405$nm with respect to the laser wavelength $\lambda=810$nm. The amplitude $A(t)$ is assumed to be zero initially, and is ramped gradually up to $E_{r}/2$ in $2.65$ms as $\textrm{exp}(kt)-1$ where $E_{r}=h^{2}/2m\lambda^{2}$ is the photon recoil energy, and $k$ is an constant determined from the ramping time. Thereafter $A(t)$ remains unchanged and the total confining potential is $V(x,t) = V_{OL}(x,t) +m \omega^{2}x^{2}/2$, where $m=1.44\times10^{-25}$kg, $\omega=2\pi\times 60$Hz is the axial angular frequency of the harmonic trap and the corresponding oscillation length is $l_b=\sqrt{\hbar/m\omega}\approx1.39\mu$m. At time $t=2.65$ms, we abruptly displace the harmonic trap through a distance $\Delta x=3\mu$m, accelerating the Bose gas in the OL.

The length of the system we used for simulations is $L=102.4l_b$, which is divided into 4096 grid points with $\delta x = 0.025l_b\approx 35$nm being the computational length unit (grid spacing). This size of the system is large enough to accommodate the dynamics of the Bose gas with $N_0=1000$ atoms (the Thomas-Fermi half length of the ground state condensate is about $R_{TF}\approx19.45l_b$), and the length unit is small enough to explore the microscopic dynamics in each well in the optical lattice. We carried out four sets of simulations of this dynamical evolution. In case I no fluctuations were added and the evolution was simply found by integrating the GP equation for the chosen initial state. In case II fluctuations were added by including long wavelength PW modes, $\phi_j(x) = e^{ik_jx}/\sqrt{L}$ where $k_j = 2\pi j/L$. The whole momentum space of the system is $\left[-40\pi/l_{b},~40\pi/l_{b}\right]$ which contains 4096 modes. The noise is just added into every other mode from $j=-190$ to $j=190$ symmetrically about 0. All the other $\xi_j$ are set to be 0.

In case III the fluctuation modes are the eigenstates of the quantum harmonic oscillator problem associated with the harmonic trap potential,
\begin{equation}
\xi(x)=\sum_n\xi_n [\frac{1}{l_b\sqrt{\pi}2^nn!}]^{1/2}e^{-\frac{x^2}{2l_b^2}}H_n(x/l_b),
\end{equation}
where $H_n$ is a Hermite polynomial and $n$ is an integer. We add the noise into $191$ SHO modes from $n=0$ to $n=190$ continuously, and then calculate the dynamics of the system in the whole coordinate space.

In case IV the modes are chosen to be
\begin{equation}\label{bog-exp}
\xi(x)=\sum_{j}u_{j}(x)\alpha_{j}-v_{j}(x)\alpha_{j}^{*},
\end{equation}
where the $u_j$ and $v_j$ functions are solutions of the projected Bogoliubov equations
\begin{equation}\label{6.12}
\left(\begin{array}{cc}
\mathcal{L} & g_{1D}N_0 \widehat{Q}\left(\psi_0(x)\right)^2\widehat{Q}\\
-g_{1D}N_0 \widehat{Q}\left(\psi^*_0(x)\right)^2\widehat{Q} & -\mathcal{L}^*
\end{array}\right)
\left(\begin{array}{c}
u_j(x)\\v_j(x)\end{array}\right) = E_j \left(\begin{array}{c}
u_j(x)\\v_j(x)\end{array}\right)
\end{equation}
with non-negative energy ($E_j>0$), where $\mathcal{L} = -\hbar^{2}\partial_{x}^{2}/2m+U_{eff}(x)$,
\begin{equation}\label{effect potential} U_{eff}(x)=V(x)+ g_{1D}N_0\widehat{Q}\left\vert\psi_0(x)\right\vert^2\widehat{Q}+ g_{1D}N_0\left\vert\psi_0(x)\right\vert^2-\mu
\end{equation}
and the projection operator $\widehat{Q}=\widehat{1}-\left\vert\psi_0\right\rangle \left\langle \psi_0\right\vert$ (in other words it's integral kernel is
\[
 Q(x,x') = \delta(x-x')-\psi_0(x)\psi_0^*(x')\qquad)
\]
projects any function onto the subspace orthogonal to $\psi_0$. The noise is added into the first 191 low-energy Bogoliubov modes ($j\in[1,~191]$). As before, the amplitudes $\alpha_j$, $\alpha_j^*$ are chosen such that $\left\langle \alpha_j^*\alpha_j\right\rangle=1/2$. In a similar way to Ref.\cite{twa2}, the functions of $x$ are expanded in a suitable basis to obtain a large matrix which is diagonalized to obtain the mode functions and excitation spectra. The real space form of $\xi(x)$ is then decided by Eq.\ref{bog-exp}.

In using different basis for the initial expansion, we always restrict our discussions at zero temperature and keep the number of fluctuation modes in the system a constant. In case II (PW basis) quantum fluctuations are added into the momentum space between $(-3.72\pi/l_b, 3.72\pi/l_b)$, where the corresponding energy space is $[0, 68\hbar\omega)$. In case III quantum fluctuations are added into the SHO modes between $[0, 190\hbar\omega]$, which means that the maximum energy is nearly three times larger than that of case II. Therefore, it is expected logically, that the added quantum noise in case III should affect the dynamical properties of the system more than that in case II. However, we will show in this paper that the result is contrary to this prediction. Note that in the time evolution of the classical field, we calculate the correlation function and the condensate density by subtracting $1/2$ virtual particle from the same modes where they were added. The groundstates which correspond to optical lattice amplitudes $A=0$ and $E_{r}/2$ are calculated numerically by evolving the GP equation in imaginary time \cite{PRE.62.7438}.

The plane waves are the simplest computationally as the field of virtual particles $\xi(x)$ is obtained using the Fast Fourier transform method. The SHO basis has the advantage that the support of each mode is weighted towards the region in which the cold atoms are likely to be found and, while the construction of $\xi(x)$ is more complicated, the modes are at least pre-defined and are relatively straightforward to generate iteratively. The SCB modes should be good approximations to the quasi-particle excitations of the system, at least for weak interactions, and are well adapted to the specifics of the potential in which the cloud moves. The calculation of $\xi(x)$ in more computationally demanding because the form of the modes has to be determined by diagonalization of a large matrix whose elements depend on the form chosen for the initial order parameter field, $\psi_0(x)$.

\begin{figure}[t]
\centering
\includegraphics[scale=0.45]{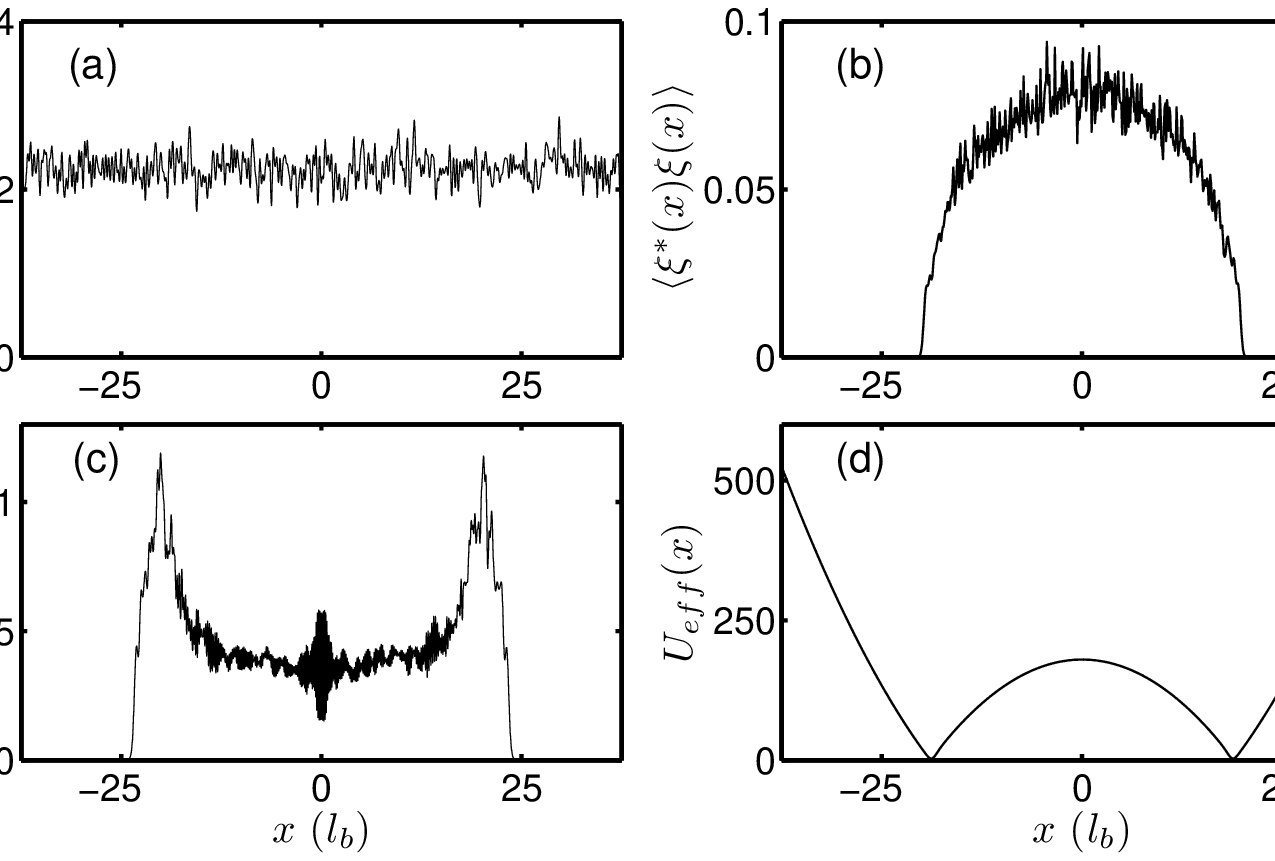}
\caption{The spatial distribution of density fluctuations, attributed respectively by adding a half quantum into each PW (a), SHO (b), and SCB mode (c). The effective potential $U_{eff}$ defined by equation \eref{effect potential} (d).}
\label{Chap5_Fig5}
\end{figure}

To determine the validity and precision of the results, different numerical methods are applied in our simulation. In case I, we obtain dynamic results by using the Crank-Nicolson method to solve the time-dependent Gross-Pitaevskii equation. In case II, a distinct numerical method, RK4IP-P \cite{PRL.94.040401,PRA.73.043617}, is utilized. Both methods are used in case III-IV to examine their equivalence. Solutions of the c.m. trajectory, phase coherence and number fluctuation in TWA, are calculated numerically using different numbers of realizations. We find that some results can be significantly different when the number of realizations is small, i.e. less than $50$. Therefore, we show all results based on 200 realizations for numerical consistency.

\section{Results and discussion}\label{disc}

In figure \ref{Chap5_Fig5}, we show the initial spatial distributions of the quantum fluctuations $\langle \xi^{*}(x)\xi(x) \rangle$ with respect to three sets of basis modes. In the PW modes, $ \langle\xi^{*}(x)\xi(x)\rangle= \frac{1}{2L} \sum_{i, j=1}^{M} \langle\xi_{i}^{*}\xi_{j}\rangle \textrm{exp}[i(k_{i}-k_{j})x]$, indicating a uniform spatial distribution of the quantum fluctuations, while the \emph{limited} number of low-energy SHO modes are dominated by the Gaussian function, which induces a localized spatial distribution of the quantum fluctuations. In contrast, the quantum fluctuations in the SCB modes give a spatial distribution with  a double peak beside the atom cloud, which coincides with the effective potential of the quasiparticles (equation \eref{effect potential}) \cite{csordas2}, as shown in figure \ref{Chap5_Fig5}(d).

\begin{figure}[t]
\centering
\includegraphics[scale=0.45]{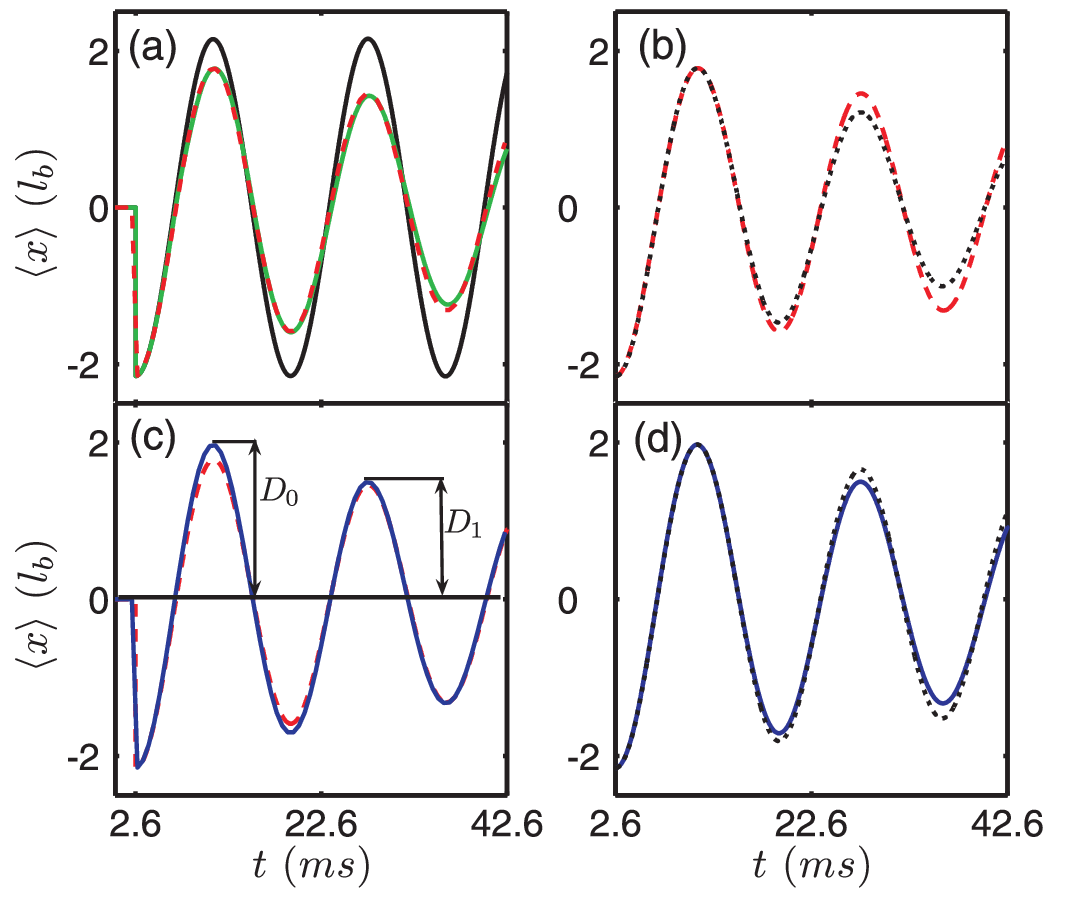}
\caption{The time evolution of the c.m. motion of the Bose gas in cases I (solid black line), II (solid green line), III (dashed red line), and IV (solid blue line). The fitting of the c.m. trajectories for cases III and IV (dotted black line in (b) and (d)) from equation \eref{cm}. The corresponding parameters: $B=0.22 \mu$m, $0.1 \mu$m and $\gamma=22.89, 10.44$ for cases III and IV, respectively , while $\Omega=2\pi\times 60$ and $A=3.0 \mu$m for both cases.}
\label{Chap5_Fig6}
\end{figure}

Qualitative insight into the properties of the quantum dynamics can be gained by using classical dissipative dynamics to compare with our quantum simulations. We model the c.m. motion as a damped harmonic oscillator \cite{PRL.94.120403, PRL.95.110403,PRA.72.011601,PRA.74.063625}
  \begin{equation}\label{5.32}
    \ddot{X}_{c.m.}+2\gamma\dot{X}_{c.m.}+\frac{k}{m^*}X_{c.m.}=0,
  \end{equation}
where $m^*$ is the effective mass and the c.m. displacement is defined by
\begin{equation}
X_{c.m.}\equiv \langle x \rangle(t)=\frac{\int_{-\infty}^{\infty}x|\Psi(x,t)|^{2}dx}{\int_{-\infty}^{\infty}|\Psi(x,t)|^{2}dx}.
\end{equation}
In underdamped case, equation \eref{5.32} have a solution
\begin{equation} \label{cm}
X_{c.m.}=-e^{-\gamma(t-t_0)}\left[A\cos\Omega(t-t_0)+B\sin\Omega(t-t_{0})\right]
\end{equation}
with $B={\gamma A}/{\Omega}$, $\Omega=\sqrt{k/m^*-\gamma^{2}}$, and $t_0$ determined by the initial phase.

\begin{figure}[t]
\centering
\includegraphics[scale=0.5]{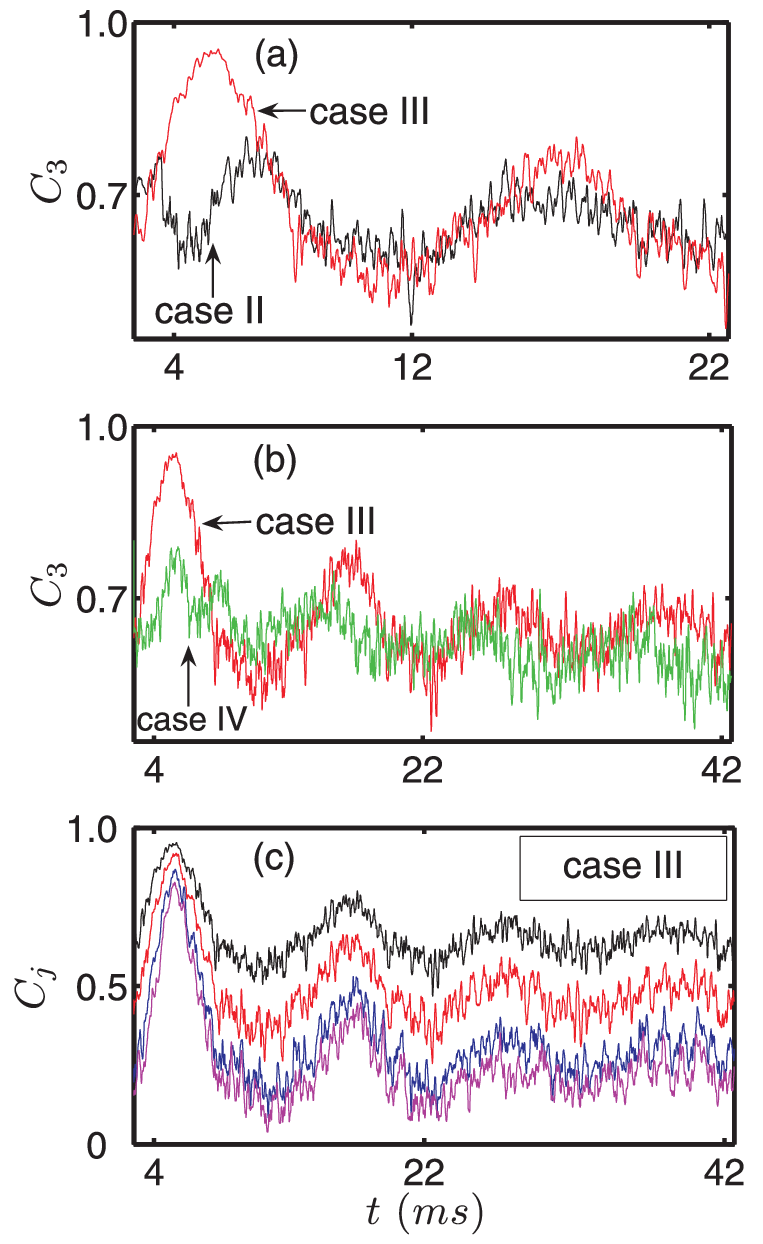}
\caption{Phase coherence $C_{3}$ in different cases (a), (b), and $C_{j}$ in case III (curves from top $C_{3}$, $C_{5}$, $C_{8}$, $C_{10}$) (c).}
\label{Chap5_Fig7}
\end{figure}

In figure \ref{Chap5_Fig6}, we show simulation results of the c.m. trajectories for case I-IV in comparison with equation \eref{cm}. In case I, there is  undamped motion of the c.m. while the visibly damped trajectories occur in presence of the quantum fluctuations in the other cases.
The damped trajectories $\langle x \rangle(t)$ cannot be fitted by equation \eref{cm}, especially at long times, indicating that the fluctuation-induced damping motion of the Bose gas cannot be depicted simply as classical damping motion. Therefore, the chosen fitting parameters satisfy the numerical calculations well in the first period but hardly characterize the subsequent behaviour of the damped motion. To compare the time-dependent damping rate in different cases, we introduce $\chi \equiv\textrm{ln}(D_{0}/D_{1})$ \cite{PRL.93.070401}, where $D_{0}$ and $D_{1}$ are the c.m. positions at $t=10.9$ms and $27.6$ms (figure \ref{Chap5_Fig6}(c)). Our calculations for cases III and IV (figure \ref{Chap5_Fig6}(c)) give $\chi=0.21$ and $0.26$ respectively, therefore indicating that, with the same number of basis states, the quantum fluctuations in SCB modes inhibit more strongly the motion of the Bose gas than in other modes. As shown in figure \ref{Chap5_Fig6}(a) the c.m. trajectories in cases II (PW basis, solid green line) and III (SHO basis, dashed red line) are nearly overlapped up to the first oscillation period, and there are only trivially small differences after that. As in our simulations quantum fluctuations in SHO basis occupy higher energy (momentum) space (about three times higher) than in PW basis, it reveals that while the identical number of quantum fluctuations are added the simulation using SHO basis is dynamically stable. This highlights the importance of the means of adding quantum fluctuations in studying different dynamical phenomena.

In figure \ref{Chap5_Fig6}(b) and \ref{Chap5_Fig6}(d), we can see that by choosing appropriate parameters for equation \eref{cm} a very good fit may be achieved with the numerical results in the first period, but the subsequent c.m. damped motion is poorly characterized. Furthermore, the fitting results from the analytical prediction \eref{cm} give a larger estimate of the c.m. damping rate than the simulation in SHO basis, but give a smaller estimate than the simulation in SCB basis. This indicates that the role of the quantum fluctuations can not be simply equivalent to a frictional force proportional to the c.m. velocity of the Bose gas; this highlights the nonlinear effect raised by the interaction between virtue particles and condensate atoms accessing into the transport of the system.

\begin{figure}[t]
\centering
\includegraphics[scale=0.50]{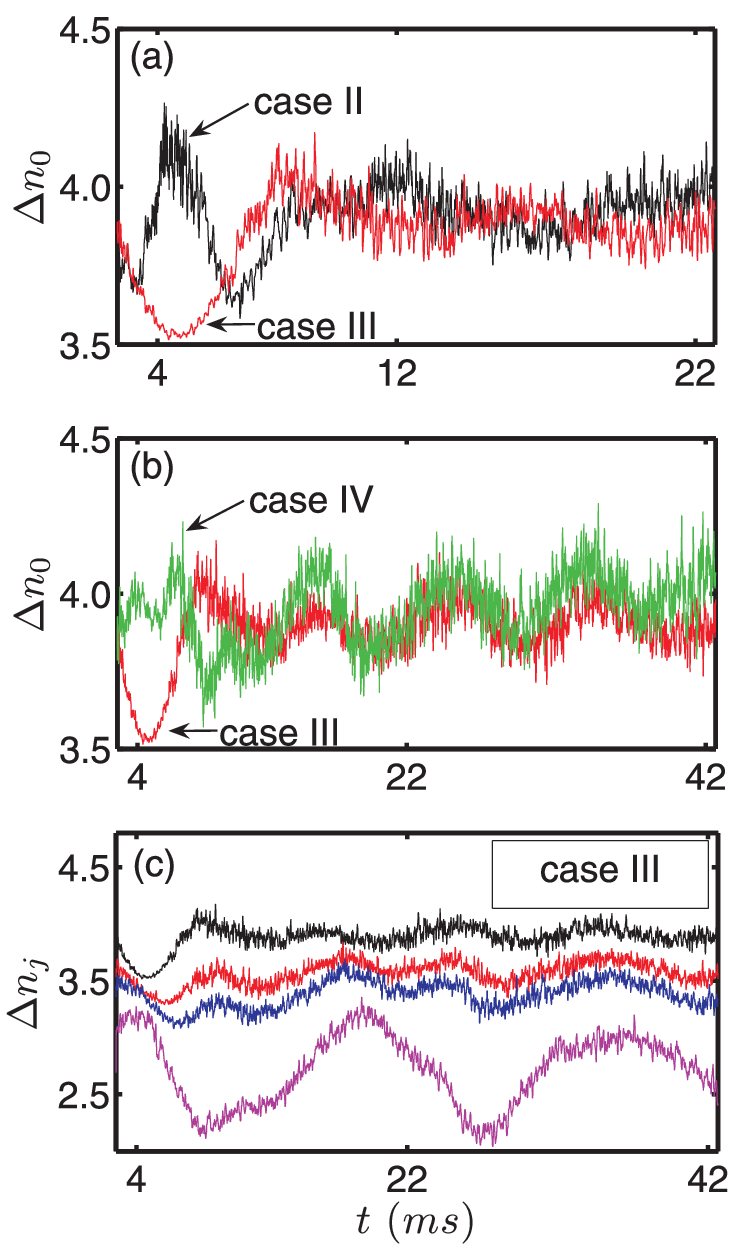}
\caption{Number fluctuation $\Delta n_{0}$ in different cases (a), (b), and $\Delta n_{j}$ in case III (curves from top $\Delta n_{0}$, $\Delta n_{10}$, $\Delta n_{20}$, $\Delta n_{40}$) (c).}
\label{Chap5_Fig8}
\end{figure}

The damped motion of the c.m. induced by quantum fluctuations suggests that there might be a loss of phase coherence, and number fluctuations in the system. In order to avoid the complications arising from the symmetrically ordered multimode field $\Psi$, we follow \cite{PRL.95.110403,PRA.72.011601,PRA.74.063625} and define the ground state operators $b_{j}$ for each individual lattice site $j$:
\begin{equation}
b_{j}(t)=\int_{j^{th}well}dx\Psi_{0}^{*}(x)\Psi(x,t),
\end{equation}
where $\Psi(x,t)$ is determined by equation \eref{5.26} and $\Psi_{0}(x)$ is the groundstate wave function in the combined harmonic trap and OL. The normalized phase coherence between the central well and its $j_{th}$ neighbor
and the atom number fluctuations in the $j_{th}$ site aref
defined as $C_{j}=|\langle\hat{b}_{0}^{\dag}\hat{b}_{j}\rangle|/\sqrt{n_{0}n_{j}}$
and $\Delta n_{j}=[\langle(\hat{b}_{j}^{\dag}\hat{b}_{j})^{2}\rangle-\langle\hat{b}_{j}^{\dag}\hat{b}_{j}\rangle^{2}]^{1/2}$. The $0_{th}$ site is located in the centre of the trap.

Figure \ref{Chap5_Fig7} shows the evolution of the dynamic phase coherence for cases II-IV ((a) and (b)) and on different spatial sites for case III ((c)) after $t=2.65ms$ when quantum fluctuations are added. We can see that $C_j$ becomes time-dependent when incorporating quantum fluctuations. In cases III and IV, within the first period of the c.m. oscillation, there is an increase in the phase coherence $C_{3}$ for $t<4.8$ms and then a decrease for $t>4.8$ms. The peak value of the phase coherence in the following periods decreases. The fact that the phase coherence never reached 1 demonstrates that the introducing of quantum fluctuations suppresses the phase coherence. Moreover, the phase coherence $C_{j}$ is correlated essentially to the damped trajectory of the c.m.. At $t=2.65$ms, the c.m. in figure \ref{Chap5_Fig6} is displaced from the centre of the trap potential, and thus $C_{3}$ reflects the properties in a region of the Bose gas where
the phase coherence is weaker. Then the c.m. of the Bose gas moves toward the centre of the trap and $C_{3}$ increases. Clearly the average value of $C_j$ is lower for larger $j$ because the first-order spatial phase coherence for 1D Bose gases tends to decay with distance, but the amplitude of the oscillations in $C_j$ are larger for larger $j$ due to the density variation across the cloud.

In figure \ref{Chap5_Fig8} we show the number fluctuation, $\Delta n_0$ for cases II and III (a), cases II and IV (b) and $\Delta n_j$ for $j=0,10,20,40$ (c). It is clear that the evolution of number fluctuation $\Delta n_0$ is anti-correlated with that of the coherence, $C_3$.

However, qualitative differences are particularly marked at short times, as depicted in figure \ref{Chap5_Fig7}(a)
and figure \ref{Chap5_Fig8}(a), indicating at least a marked dependence of the ``equilibration time'' (i.e. the period for the three different basis sets we used to get similar results) for such simulations on the choice of basis set. This is most marked for the PW basis in which the short-time oscillations in the phase coherence are quite different to the others and only approach them at times rather longer than the period of the dipole oscillations. \ref{Chap5_Fig7}(b) and \ref{Chap5_Fig8}(b) show that the short-time variation of $C_{3}$ and $\Delta n_{0}$ based on the SCB modes is more complicated than the other two modes. This might originate from the greater complexity of the initial distribution of the quantum fluctuations in the SCB modes.  Comparing the phase coherence for cases III and IV, shown in figure \ref{Chap5_Fig7}(b), we find that the time average of $C_{3}$ from $2.65$ms to $42.65$ms for case IV is $0.63$, $5.21\%$ smaller than that for case III ($0.66$). The corresponding mean value for the number fluctuation in the time interval for case IV is $3.95$, a little higher than $3.88$ for case III. This indicates that, with the same number of basis modes, the quantum fluctuations in the SCB modes cause stronger phase decoherence than those in the PW and SHO modes.

\begin{figure}[t]
\centering
\includegraphics[scale=0.5]{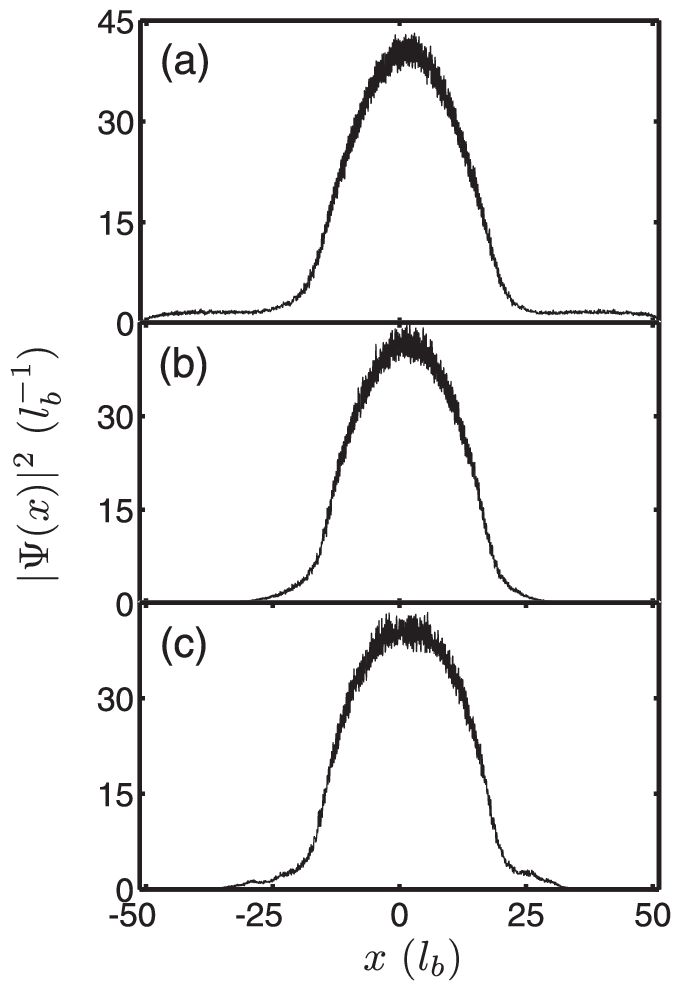}
\caption{Density distribution at $t=8.65$ms in cases II (a), III (b) and IV (c). This shows the density averaged over 200 realizations of the random initial conditions.}
\label{Chap5_Fig10}
\end{figure}

Interestingly, we also find that, during the transport process, some atoms are kicked out from the core region of the atom cloud in case II (figure \ref{Chap5_Fig10}(a)) while the quantum fluctuations only break the inner configuration of the condensate slightly in cases III (figure \ref{Chap5_Fig10}(b)) and IV (figure \ref{Chap5_Fig10}(c)). In all cases the data shown is the result of averaging over 200 realizations of the random initial conditions. In the experiment \cite{PRL.94.120403}, they did not observe a significant difference in the time of flight width between atoms that undergo damped harmonic motion and those that are unexcited but held for an equal time.

Apparently we can strictly only say that the classical field dynamics with different initial states based on
the expansion of quantum fluctuations in three basis sets produces different results, but it is natural to assume that the simulations using the SCB modes have more information about the specific physics of the system built into them and should be more reliable. We then note that for some properties, the gain in using the SCB modes compared to the SHO modes is small given the considerable increase in computational complexity. However we note that in the simulation that uses PW modes for trapped BECs, a long equilibration time should be allowed before any non-trivial dynamical processes are allowed to occur in the simulation to avoid spurious transient effects.

\section{Conclusion}\label{conc}
In conclusion, we have studied the effect of the different choices of basis set for the inclusion of quantum fluctuations on a 1D Bose gas in an OL using the TWA. Specifically we have used PW, SHO and the SCB modes to decompose the quantum fluctuations and examine the dynamics of the system in the whole coordinate space.  The difference in the predictions of the phase coherence, number fluctuations and density variations at short time indicates that the choice of the basis set for the initial expansion has a substantial influence on the qualitative features of the transport. The use of the SCB modes, which gives greater phase decoherence and stronger number fluctuations on the damping dynamics than the other two choices, incorporates more of the underlying physics of the problem and presumably leads to the most reliable results. For the trapped condensate, the SHO basis give results that are qualitatively similar to those from the SCB modes, at rather less computational effort, although some of the detail of the fluctuations at short time are distinct. It's therefore clear that when quantum fluctuations are included it is important to select the basis set for the initial state of the system that best suits the particular system to be studied.
\section*{ACKNOWLEDGMENTS}
We thank R. G. Scott and Carlos Lobo for many useful discussions, and S. Chen for suggestions in the numerical work. We also acknowledge support from the EPSRC.

\section*{References}
\bibliographystyle{unsrt}
\bibliography{ThesisEx}

\end{document}